%% file: main.tex
\documentclass[sigconf]{acmart}

\AtBeginDocument{%
  }

\setcopyright{acmlicensed}
\copyrightyear{2025}
\acmYear{2025}
\acmConference[WWW '25] {Proceedings of the ACM Web Conference 2025}{April 28--May 2, 2025}{Sydney, NSW, Australia.}
\acmBooktitle{Proceedings of the ACM Web Conference 2025 (WWW '25), April 28--May 2, 2025, Sydney, NSW, Australia}
\acmISBN{979-8-4007-1274-6/25/04}

\acmDOI{10.1145/3696410.3714576}

\acmISBN{979-8-4007-1274-6/25/04}

\input{packages}
\input{commands}

\graphicspath{{./figures/}}

\begin{document}

\title{Towards Collaborative Anti-Money Laundering Among Financial Institutions}

\author{Zhihua Tian}
\affiliation{%
\institution{Zhejiang University}
\city{Hangzhou}
\country{China}}
\email{zhihuat@zju.edu.cn}

\author{Yuan Ding}
\affiliation{%
\institution{Zhejiang University}
\city{Hangzhou}
\country{China}}
\email{dy1ant@zju.edu.cn}

\author{Wenjie Qu}
\affiliation{
\institution{National University of Singapore}
\country{Singapore}}
\email{wenjiequ@u.nus.edu}

\author{Xiang Yu}
\affiliation{%
\institution{Alibaba Group}
\city{Hangzhou}
\country{China}}
\email{shaseng.yx@antgroup.com}

\author{Enchao Gong}
\affiliation{%
\institution{Alibaba Group}
\city{Hangzhou}
\country{China}}
\email{enchao.gec@antgroup.com}


\author{Jian Liu}
\affiliation{%
\institution{Zhejiang University}
\city{Hangzhou}
\country{China}}
\email{liujian2411@zju.edu.cn}
\authornote{Corresponding author}

\author{Kui Ren}
\affiliation{%
\institution{Zhejiang University}
\city{Hangzhou}
\country{China}}
\email{kuiren@zju.edu.cn}

\renewcommand{\shortauthors}{Zhihua Tian et al.}

\begin{abstract}
Money laundering is the process that intends to legalize the income derived from illicit activities, thus facilitating their entry into the monetary flow of the economy without jeopardizing their source. It is crucial to identify such activities accurately and reliably in order to enforce anti-money laundering (AML).

Despite considerable efforts to AML, a large number of such activities still go undetected. Rule-based methods were first introduced and are still widely used in current detection systems. With the rise of machine learning, graph-based learning methods have gained prominence in detecting illicit accounts through the analysis of money transfer graphs. Nevertheless, these methods generally assume that the transaction graph is centralized, whereas in practice, money laundering activities usually span multiple financial institutions. Due to regulatory, legal, commercial, and customer privacy concerns, institutions tend not to share data, restricting their utility in practical usage.
In this paper, we propose the \textit{first} algorithm that supports performing AML over multiple institutions while protecting the security and privacy of local data.

To evaluate, we construct Alipay-ECB, a real-world dataset comprising digital transactions from Alipay, the world's largest mobile payment platform, alongside transactions from E-Commerce Bank (ECB). The dataset includes over 200 million accounts and 300 million transactions, covering both intra-institution transactions and those between Alipay and ECB. This makes it the largest real-world transaction graph available for analysis. 
The experimental results demonstrate that our methods can effectively identify cross-institution money laundering subgroups. Additionally, experiments on synthetic datasets also demonstrate that our method is efficient, requiring only a few minutes on datasets with millions of transactions.  Our code and dataset are available on~\url{https://github.com/zhihuat/Collaborative-AML}.
\end{abstract}



\maketitle

\input{sections/introduction}

\input{sections/preliminaries}

\input{sections/problem_statement}

\input{sections/methods}

\input{sections/experiments}

\input{sections/related_works}

\input{sections/conclusion}

\begin{acks}
This work is sponsored in part by National Key Research and Development Program of China (2023YFB2704000) and CCF-AFSG research fund.
\end{acks}

\bibliographystyle{ACM-Reference-Format}
\balance
\bibliography{main}

\appendix
\newpage
\input{sections/appendix}

\end{document}

%% file: packages.tex
\usepackage{tikz}
\usepackage{amsmath}
\usepackage{mathrsfs}
\usepackage{xcolor}
\usepackage{dsfont}
\usepackage{numprint}
\usepackage{subfig}
\usepackage{caption}
\usepackage[export]{adjustbox}

\usepackage{url}
\usepackage{setspace}
\usepackage{tabularx}  

\usepackage{indentfirst}
\usepackage[ruled,lined,linesnumbered]{algorithm2e}  
\usepackage{algorithmic}
\usepackage{tikz}
\usetikzlibrary{calc,fit}

\colorlet{pink}{red!40}
\colorlet{blue}{cyan!60}

\usepackage{booktabs}
\usepackage{nicematrix}
\usepackage{graphicx}
\usepackage{pgfplots}
\pgfplotsset{compat=1.17}

\usepackage{color}
\usepackage{verbatim}
\usepackage{textcomp}

\usepackage{bbold}

\usepackage[utf8]{inputenc}
\usepackage{array}
\usepackage{makecell}
\usepackage{wrapfig}

\usepackage{multirow}
\usepackage{multicol}

\usepackage{soul}
\usepackage{balance}

\usepackage{natbib}
 \usepackage[normalem]{ulem}

%% file: commands.tex
\newcommand{\Paragraph}[1]{\noindent{\textbf{#1}}}

\newcommand{\lsh}{\mathit{lsh}}

\newcommand{\SSS}{\mathcal{S}}

\newcommand{\Cli}{\mathcal{P}}
\newcommand{\bal}{\textit{AMLSim\_bal}}
\newcommand{\unb}{\textit{AMLSim\_unb}}
\newcommand{\hi}{\textit{AMLWorld\_HI}}
\newcommand{\li}{\textit{AMLWorld\_LI}}

\newtheorem{theorem}{Theorem}
\newtheorem{innercustomthm}{Theorem}
\newenvironment{customthm}[1]
  {\renewcommand\theinnercustomthm{#1}\innercustomthm}
  {\endinnercustomthm}

%% file: sections/introduction.tex
\section{Introduction}
\begin{figure}[t]
\centering
\subfloat[]{
    \label{fig: topology1}
    \includegraphics[width=0.48\linewidth]{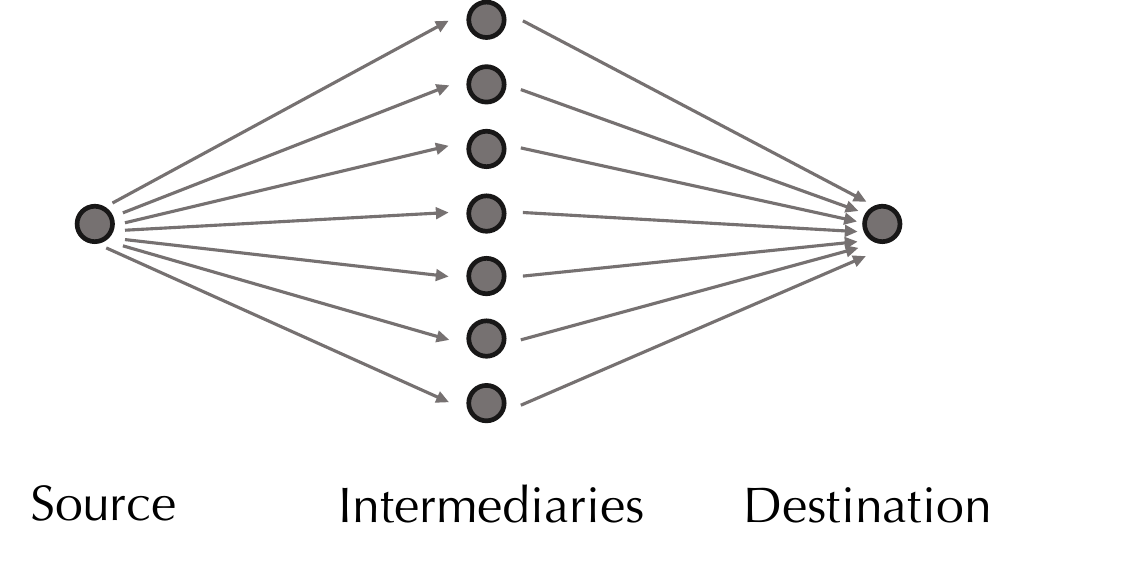}}
\subfloat[]{
    \label{fig: topology2}
    \includegraphics[width=0.48\linewidth]{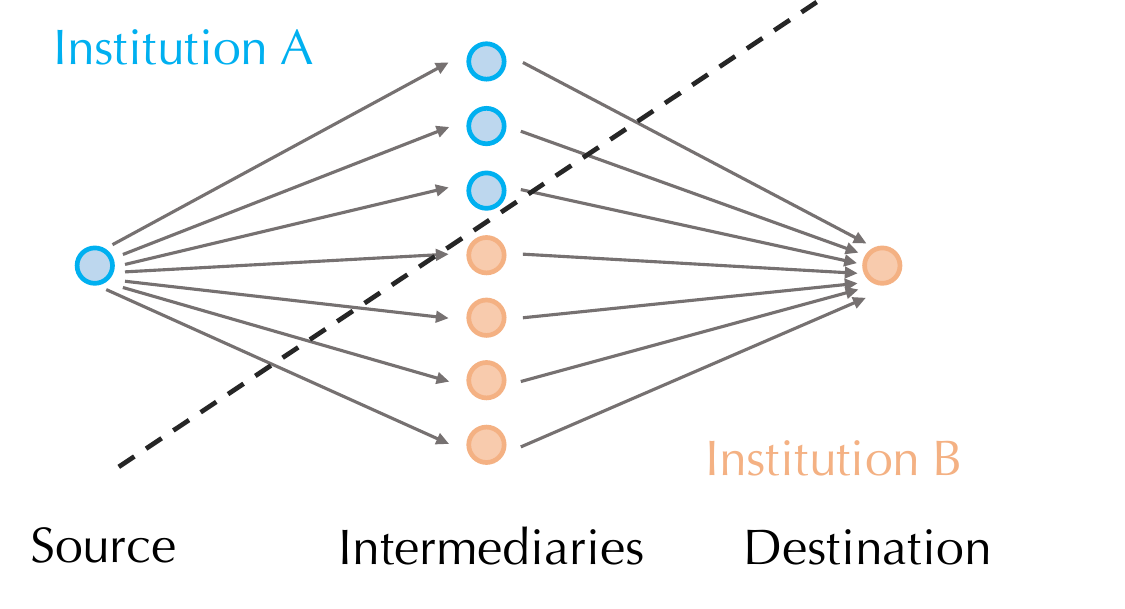}}
\caption{(a) Scatter-gather pattern  money laundering; (b) Scatter-gather distributed across two institutions.}
\end{figure}

Money laundering is a process that attempts to conceal or disguise the origins of dirty money derived from illicit activities, making it appear as if the funds have been obtained through legitimate means~\cite{aml}.
It typically consists of three primary steps: a \textit{placement} step first introduces the dirty money into existing financial systems; 
a \textit{layering} step then carries out complex transactions to hide the source of the funds; 
and a \textit{integration} step withdraws the fund from a destination bank account before using it for legitimate activities~\cite{unitednations}.
The transaction relationship of accounts can be represented as a graph, where an individual account is denoted as a node, and transactions between two accounts are denoted as edges. Due to the distinctive nature of money laundering activities, the transaction graph associated with money launderers exhibits a unique pattern known as \textbf{scatter-gather}~\cite{michalak2011graph, altman2024realistic, egressy2024provably}, as illustrated in Fig.~\ref{fig: topology1}.

It is the responsibility of financial institutions to conduct {\em anti-money laundering} (AML): 
diligently monitor transactions, take necessary actions like shutting down or imposing restrictions on suspicious accounts, and promptly report any suspicious activities through to law enforcement agencies. 
To detect money laundering activities, a common idea is to identify the ultimate beneficiary, which refers to the individual or entity that ultimately receives the funds, even if those funds have been obscured through multiple layers of transactions~\cite{aml}. 
To achieve that, a simple approach is to calculate the ratio to which funds in one account originate from another account~\cite{michalak2011graph}. 
If the ratio exceeds a predefined threshold, it indicates a potential association between the two accounts, raising suspicions of money laundering activities with one account being the source and the other the destination.

However, money laundering has evolved into a highly sophisticated process, spanning across multiple financial institutions s.t. the subgraph within one institution appears to be normal (Fig.~\ref{fig: topology2}). 
As a result, relying solely on the transaction graph within a single institution for AML is no longer sufficient. 
A straightforward solution is to combine the transaction graphs from multiple institutions. However, due to regulatory, legal, commercial, and customer privacy concerns, institutions tend not to share data.


\Paragraph{Our contribution.}
In this paper, we make the \textit{first} step towards collaborative AML,
which allows multiple institutions to jointly conduct AML without exposing their individual transaction graphs.

Our primary contribution lies in the introduction of a novel algorithm for scatter-gather subgraph mining, specifically tailored to suit the collaborative setting. 
In more detail, this algorithm first employs a breadth-first search (BFS) approach for each node to identify a set of cross-institution transactions associated with that node, which can be either scattered from or gathered towards the node.
If two nodes, belonging to different institutions, share the same set of cross-institution transactions, it indicates a potential scatter-gather relationship within a money laundering subgraph, with one node being the source and the other being the destination.
Building upon this observation, the algorithm considers two institutions, denoted by $\Cli_A$ and $\Cli_B$, and iterates through their respective nodes ($\{N_1^A, N_2^A, \ldots, N_n^A\}$ and $\{N_1^B, N_2^B, \ldots, N_n^B\}$) to identify the sets of cross-institution transactions: $\SSS^A = \{S_1^A, S_2^A, \ldots, S_n^A\}$ and $\SSS^B = \{S_1^B, S_2^B, \ldots, S_n^B\}$, where e.g., $S_i^A$ is the set of cross-institution transactions associated with node $N_i^A$. 
If two sets $S_i^A$ and $S_j^B$ exhibit a high degree of similarity, it suggests that $N_i^A$ and $N_j^B$ are potentially involved in scatter-gather activities within a money laundering subgraph.

This approach requires $\Cli_A$ and $\Cli_B$ to exchange $\SSS^A$ and $\SSS^B$, and measure the similarity between each pair (e.g., $S_i^A$ and $S_j^B$). This is costly in terms of both communication and computation.
To solve the problem, we use locality-sensitive hashing (LSH)~\cite{lsh} and Bloom filter~\cite{bloomfilter} to minimize the amount of information to be exchanged between $\Cli_A$ and $\Cli_B$. 
LSH enables the estimation of similarity between two sets by comparing the minimum hash values of their elements. Combined with Bloom filters, the approach transforms pairwise comparisons into a process of testing the presence of an element within a Bloom filter.  The Bloom filter is memory-efficient, and this testing process is computationally efficient.

Specifically, an LSH is computed for each set, resulting in $\{\lsh_1^A, \lsh_2^A, \allowbreak\ldots, \lsh_n^A\}$ and $\{\lsh_1^B, \lsh_2^B, \ldots, \lsh_n^B\}$.
Notice that $\lsh_i^A=\lsh_j^B$ if $S_i^A$ and $S_j^B$ exhibit a high degree of similarity.
Next, one institution, say $\Cli_A$, inserts $\{\lsh_1^A, \lsh_2^A, \ldots, \lsh_n^A\}$ into a bloom filter $BF_A$, and transfers $BF_A$ to $\Cli_B$;
$\Cli_B$ iterates through $\{\lsh_1^B, \lsh_2^B, \ldots, \lsh_n^B\}$ to check if each $\lsh^B$ is present in $BF_A$.
If $\lsh_j^B$ is found in $BF_A$, $\Cli_B$ learns that $N_j$ is one end node in the scatter-gather activity. 
At this stage, $\Cli_B$ reveals the corresponding $\lsh_j^B$ to $\Cli_A$, enabling $\Cli_A$ to identify the other end node in the scatter-gather activity.
By leveraging this optimization,  the communication overhead is significantly reduced as it only requires the transfer of a bloom filter. 
Moreover, by comparing against a bloom filter, the computational complexity is reduced to $O(n)$, rather than $O(n^2)$ when comparing each pair individually.

To evaluate whether our methods can detect money laundering activities across multiple institutions in a real-world setting, we construct Alipay-ECB, a multi-institution transaction dataset that includes digital currency transactions from Alipay and E-Commerce Bank (ECB) users. The dataset contains over 200 million accounts and 300 million transactions. To the best of our knowledge, it is the largest real-world transaction dataset available. 

By analyzing the dataset, we find that money laundering groups possess a much more intricate structure in real-world settings, encompassing multiple simple patterns such as fan-in, fan-out, cycles, random, and bipartite, etc. However, our method can effectively identify money laundering subgroups. Experiments on synthetic datasets also demonstrate our methods can effectively and efficiently identify money laundering subgroups.

%% file: sections/preliminaries.tex
\section{Preliminaries}
This section provides the necessary background and preliminaries for understanding this paper. The frequently used notations
are presented in Table~\ref{tab:notations}.

\subsection{ Scatter-Gather Mining}
\label{ssec: MLSD}


\begin{figure}
\begin{center}
\centerline{
    \includegraphics[width=0.8\linewidth]{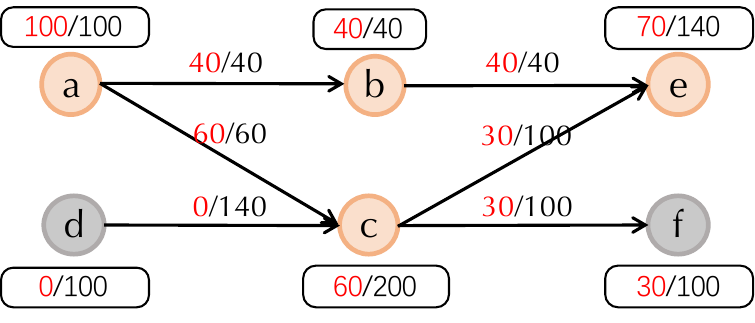}}
\caption{Illustration of centralized scatter-gather mining. Consider $a$ as the source and designate all the funds flowing out from $a$ as illicit money, visually represented in \textcolor{red}{red}. The enclosed numbers within boxes indicate the money possessed by the nodes, while numbers above the line represent the money involved in a transaction.
The orange nodes represent the detected money laundering nodes, which have an illicit funds ratio greater than 0.3.}
\label{fig: example4sgd}
\end{center}
\end{figure}

In order to detect money laundering transaction subgraphs of scatter-gather patterns, a simple approach is to examine cases where a significant amount of money flows out of one account and gets aggregated in other accounts~\cite{michalak2011graph}. We refer to this method as \textit{Centralized Scatter-Gather Mining}.
To illustrate, let's consider an example where there's a node $j$ that receives 80\% of the money flowing out from node $i$.  In this case, it's possible that both nodes $i$ and $j$, along with the nodes in-between, are involved in a potential money laundering activity, with $i$ acting as the source within the subgraph, and $j$ serving as the destination.

To determine how much of the money received by node $j$ comes from node $i$, the method utilizes a tracking mechanism based on the transaction graph. This involves marking the outflow money from node $i$ as suspected money and tracing their movement within the graph. When node $i$ sends money to another node $v$, the marked money is transferred to $v$. Similarly, if node $v$ subsequently sends money to node $j$, the marked money is also transferred to node $j$.
In the context of the method, two principles govern the flow of marked money in downstream nodes, considering that money is divisible. Denote $M_{in}^j, M_{out}^j$ as total inflow and outflow of node $j$ separately, and $m_{in}^j, m_{out}^j$ as marked inflow and outflow included in $M_{in}^j, M_{out}^j$ that satisfy $m_{in}^j\leq M_{in}^j,\ m_{out}^j\leq M_{out}^j$.

We have the following principles to calculate $m_{in}^j$:
\begin{enumerate}
\item For a node whose inflow money involves marked money, if $M_{in}^j>M_{out}^j$, then $m_{out}^j = m_{in}^j \frac{ M_{out}^j}{ M_{in}^j}$; if $M_{in}^j\leq M_{out}^j$, then $m_{out}^j = m_{in}^j$.
\item The marked inflow money of a node is the sum of marked money received from other nodes.
\end{enumerate}
After getting the value of $m_{in}^j$, we calculate the ratio of inflow money from $i$ to $j$ as $r_{ij} = \frac{m_{in}^j}{m_{out}^i}$.

Figure~\ref{fig: example4sgd} illustrates an example of applying the method by considering node $a$ as the source node and discovering the scatter-gather pattern it is involved in. By setting the threshold to $40\%$, we identify three suspected money laundering nodes $b$, $c$, and $e$, which contain $40\%$, $60\%$ and $70\%$ of marked money, respectively.

\subsection{MinHash}
\label{ssec: minhash}
MinHash~\cite{minhash} is a technique to estimate how similar two sets are, where the similarity is defined in terms of the Jaccard similarity coefficient. Specifically, let $A$ and $B$ are two sets. The Jaccard index is defined to be the ratio of the number of elements of their intersection and the number of elements of their union: 
\begin{equation}
    J(A, B)=\frac{|A \cap B|}{|A \cup B|}.
\end{equation}
Let $H$ denote the minhash function that maps a set to a real number; it has the property  
\begin{equation}
    Pr[H(A)=H(B)] = J(A, B).
\end{equation}
That is, the probability that $H(A) = H(B)$ is true is equal to the similarity $J(A, B)$.

The details of the MinHash algorithm is following:
Given a hash function $h$ that maps the members of a set $U$ to real numbers, and \textit{perm} which is a random permutation of the elements of $U$. For any set $S \subset U$, 
$H$ is defined as the minimum value of $h(perm(x))$, i.e.,

\begin{equation}
    \label{equ: minhash}
    H(S) := min\ h(perm(x)).
\end{equation}

Let $r$ be a random variable that is 1 when $H(A)=H(B)$ and 0 otherwise, $r$ is the unbiased estimator of $J(A, B)$, i.e., $E(r) = J(A, B)$. 

The MinHash scheme reduces this variance by averaging together several variables constructed in the same way, such as by applying multiple hash functions.
To estimate $J(A, B)$, let $n$ be the number of hash functions for which $H(A) = H(B)$, $\frac{n}{K}$ is the estimate, where $K$ is the total number of hash functions used. This estimate is the average of $K$ random variables $r$s, each of which is the unbiased estimator of $J(A, B)$. Hence, the average is also unbiased. By standard deviation for sums of the variables, the similarity estimation error is $\mathcal{O}(1/\sqrt{K})$.

\subsection{Bloom filter}
\label{ssec: bloomfilter}
A Bloom filter~\cite{bloomfilter} is a memory-efficient data structure that is used to test whether an element is present in a set. The price paid for the efficiency is that Bloom filter is a probabilistic data structure: It tells us that the element either \textit{definitely} is not in the set or \textit{may be} in the set. In other words, false positive matches are possible, but false negatives are not.

A Bloom filter is an array of $m$ bits with all positions set to $0$ when it is empty. There are also $k$ hash functions, each of which maps or hashes each element in a set to one of the $m$ positions uniformly. To \textit{add} an element, we simply feed it to each of the $k$ hash functions to get $k$ array positions and set the bits at all these positions to 1.
To \textit{query} an element (test whether it is in the set), hash it using the identical $k$ hash functions to get $k$ array positions. If any of the $k$ positions are $0$, the element is \textit{definitely} not in the set. If all are $1$, then the element is either in the set or the bits were set to $1$ when inserting other elements by chance, resulting in a false positive.
The false positive error $\epsilon$, the size of Bloom filter $m$, and the number of hash functions $k$ are related in the following way:
\begin{equation}
    k = -\log_2\epsilon,\ m = -\frac{n\ln\epsilon}{(\ln2)^2}, \text{and}\ \epsilon=(1-e^{-\frac{kn}{m}})^k
\end{equation}
With $m$ increases, the false positive probability $\epsilon$ decreases.

%% file: sections/problem_statement.tex
\section{Problem Statement}
\label{sec: problem_statement}

\begin{figure*}[t]
\begin{center}
\centerline{
    \includegraphics[width=0.85\linewidth]{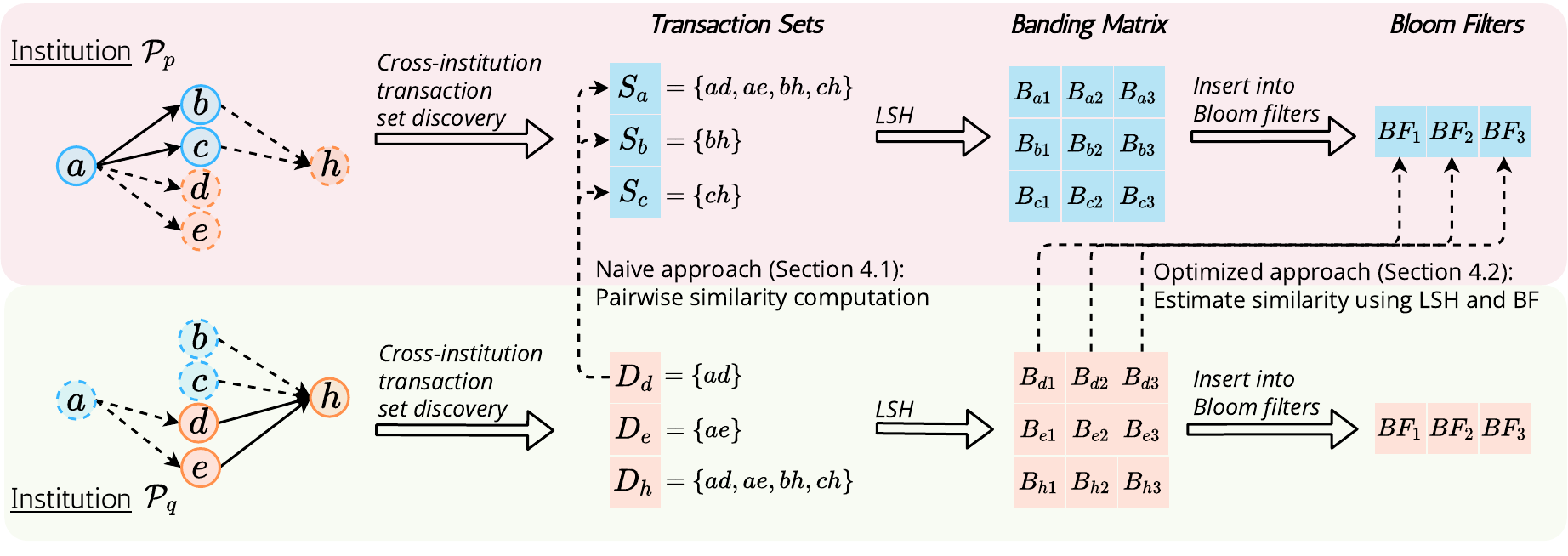}}
\caption{Workflow of CSGM. The dotted lines on the graphs indicate cross-institution transactions.}
\label{fig:workflow}
\end{center}
\end{figure*}

Let $\mathcal{G=(V,E,X)}$ be a money transaction graph, where $\mathcal{V}$ is the vertex set represents accounts, $\mathcal{E}$ is the edge set represents transactions, and $\mathcal{X} \in \mathbb{R}^d$ is the feature matrix of all edges. 
An edge $(i, j) \in \mathcal{E}$ indicates that the account $i$ transfers money to $j$ and the corresponding $\mathbf{x}\in\mathcal{X}$ indicates the attributes of the transaction, such as the amount of money, the time, to name a few. In this paper, we mainly focus on two attributes: the amount of money and whether the transaction is an external transaction, denoted as $a$ and $c$ separately. Specifically, $\mathbf{x} = [a, c]^\top$. For ease of presentation, we denote $\mathbf{x}_{i\rightarrow j}$ the attributes for the transaction from $i$ to $j$.

In our setting of collaborative learning, we consider two institutions $\Cli_A$ and $\Cli_B$; each holds a subgraph $\mathcal{G}_A = (\mathcal{V}_A,\mathcal{E}_A,\mathcal{X}_A)$ and $\mathcal{G}_B = (\mathcal{V}_B,\mathcal{E}_B,\mathcal{X}_B)$, where $\mathcal{V}_i,\mathcal{E}_i,\mathcal{X}_i$ are subsets of $\mathcal{V},\mathcal{E},\mathcal{X}$, separately.
In the rest of the paper, we use the notations $p$ and $q$ to denote the indices of the two institutions. Specifically, \(\Cli_p\) refers to one institution and \(\Cli_q\) to the other.

To comply with Know Your Customer (KYC) standards~\cite{kyc}, financial institutions are required to gather basic information about both the initiator and recipient of each transaction. This rule remains applicable even when accounts are held across different institutions. Based on this requirement, we assume an overlap between $\mathcal{V}_A$ and $\mathcal{V}_B$. The overlapping nodes represent accounts involved in cross-institution transactions between $\Cli_A$ and $\Cli_B$.

We further assume that the overlapping accounts are recorded with identical identifiers by both institutions. This identification can be performed privately through multi-party private set intersection methods~\cite{kolesnikov2017practical}, which is orthogonal to our paper.

Given the above setting, we aim to discover money laundering groups of typologies presented in figure~\ref{fig: topology1} based on two subgraphs
$\mathcal{V}_A$ and $\mathcal{V}_B$.


%% file: sections/methods.tex
\section{Methods}

In this section, we present in detail how our collaborative AML algorithm, named collaborative scatter-gather mining (CSGM), is designed. We begin by transforming the centralized scatter-gather mining method into the one that can be applied to two subgraphs as defined in Section~\ref{sec: problem_statement} owned by different institutions.
The method enables the detection of money laundering nodes distributed across multiple institutions, particularly when the source and destination nodes belong to different institutions.

We further enhance the method by making use of Locality-sensitive hashing (LSH)~\cite{lsh} and Bloom filter~\cite{bloomfilter} to minimize communication costs and improve efficiency. Figure~\ref{fig:workflow} presents the workflow of CSGM.

\subsection{Collaborative Scatter-Gather Mining}
In the scatter-gather pattern of money laundering,  money is transferred from a source to a destination through multiple transactions involving many adversarial middle nodes. 
When the source and the destination are located in different institutions, it implies that money laundering activities transfer money to another institution via cross-institution transactions, as shown in Figure~\ref{fig: topology2}.

The key idea behind our method is that the set of cross-institution transactions scattered from the source is identical to the set of cross-institution transactions gathered at the destination when the source and destination are involved in the same money laundering subgraph. Therefore, by comparing the sets of transactions identified by both institutions, we can effectively detect money laundering subgraphs in which both source and destination are implicated.




Specifically, let $\mathcal{S}_p \leftarrow [S_i \mid i \in \mathcal{V}_p]$ and $\mathcal{D}_p \leftarrow [D_i \mid i \in \mathcal{V}_p]$ for \(p \in \{A, B\}\) denote the sets of all cross-institution transactions associated with $\Cli_p$, where $S_i$ and $D_i$ represent the sets obtained through scattering from or gathering to node $i$, respectively. $\Cli_p$ transmits both $\mathcal{S}_p$ and $\mathcal{D}_p$ to institution $\Cli_q$, ensuring that both $\Cli_A$ and $\Cli_B$ possess all relevant sets.
By independently comparing the similarity between any two sets $S_i \in \mathcal{S}_p$ and $D_j \in \mathcal{D}_q$, each institution can identify sources or destinations involved in money laundering activities. Specifically, $\Cli_p$ can detect sources by comparing sets from $\mathcal{S}_p$ with those from $\mathcal{D}_q$, and similarly, identify destinations by comparing sets from $\mathcal{D}_p$ with those from $\mathcal{S}_q$. Note that we filter out the discovered sets of small size (setting the threshold to 4-7 in our experiments), considering that money laundering groups are typically huge to conceal substantial amounts of money.
Once all suspicious sources and destinations are identified, intermediate nodes can be readily located by tracing the transactions that are scattered from sources or gathered to destinations within the local subgraph.

\Paragraph{Cross-institution Transaction Set Discovery.}
To find the set of cross-institution transactions, each institution employs the BFS approach for each node to find transactions scattered or gathered from the node and determine if they are cross-institution transactions. 
Specifically, it starts from a specific node and loops all neighbor nodes to identify cross-institution transactions originating from the node until either all relevant transactions are found or the maximum depth is reached.
Let $\mathcal{F}$ represent the algorithm, and we denote the discovery process as $S_i\leftarrow \mathcal{F}(i, \mathcal{G}, T)$.
Here, $S_i$ is the set of cross-institution transactions scattered from node $i$, $\mathcal{G}$ is the local transaction graph, and $T$ denotes the maximum depth allowed.
When aiming to discover the gathered transaction sets,
we can simply transform $\mathcal{G}$ into a new graph $\mathcal{G}'$ with the inverse direction. Specifically, $\mathcal{G}'= (\mathcal{V}, \mathcal{E}', \mathcal{X})$, where $\mathcal{E}' = \{(j, i)| \left(i, j\right)\in \mathcal{E}\}$. By performing the same algorithm on $\mathcal{G}'$, we construct the reversed transaction set as $D_i \leftarrow \mathcal{F}(i, \mathcal{G}', T)$. 
Algorithm~\ref{alg:set_discovery} in Appendix presents the procedure. 

\subsection{Optimization for Distributed Scatter-Gather Mining}
Applying the distributed scatter-gather mining algorithm directly is both communication- and computation-intensive, as it requires institutions to exchange multiple transaction sets ($\mathcal{O}(n)$, where $n$ is the number of nodes) and perform pairwise comparisons among them, which is $\mathcal{O}(n^2)$. To address the challenge, we propose an optimized algorithm using LSH~\cite{lsh} and Bloom filters~\cite{bloomfilter}. LSH enables the estimation of similarity between two sets by comparing the minimum hash values of their elements, and by inserting the results of all sets (either from $\mathcal{S}$ or $\mathcal{D}$) into a bloom filter, we transform pairwise comparisons into a more efficient process of testing whether an element exists within the Bloom filter.

Specifically, institution $\Cli_p$ first performs LSH on all sets. The results are then inserted into $K$ Bloom filters, where $K$ is determined by the length of the LSH value. 
The Bloom filters are then shared with another institution, $\Cli_q$. By querying the Bloom filter with the LSH of $\Cli_q$'s local set, which is likely to match those of other sets with high similarity, $\Cli_q$ can efficiently detect the existence of a similar set, thereby determining whether the corresponding node is involved in potential money laundering activities.  As it requires only the transfer of Bloom filters, the optimization significantly reduces communication overhead, Moreover, the computational complexity is reduced to $\mathcal{O}(Kn), k<<n$, as opposed to $\mathcal{O}(n^2)$ when performing pairwise comparisons.

Next, we provide a detailed explanation of how sets are inserted into Bloom filters. We then introduce two methods, namely Probability-Based Similar Set Detection and Similarity-Based Similar Set Detection, to detect similar sets using Bloom filters.

\begin{figure}[t]
\begin{center}
\centerline{
    \includegraphics[width=\linewidth]{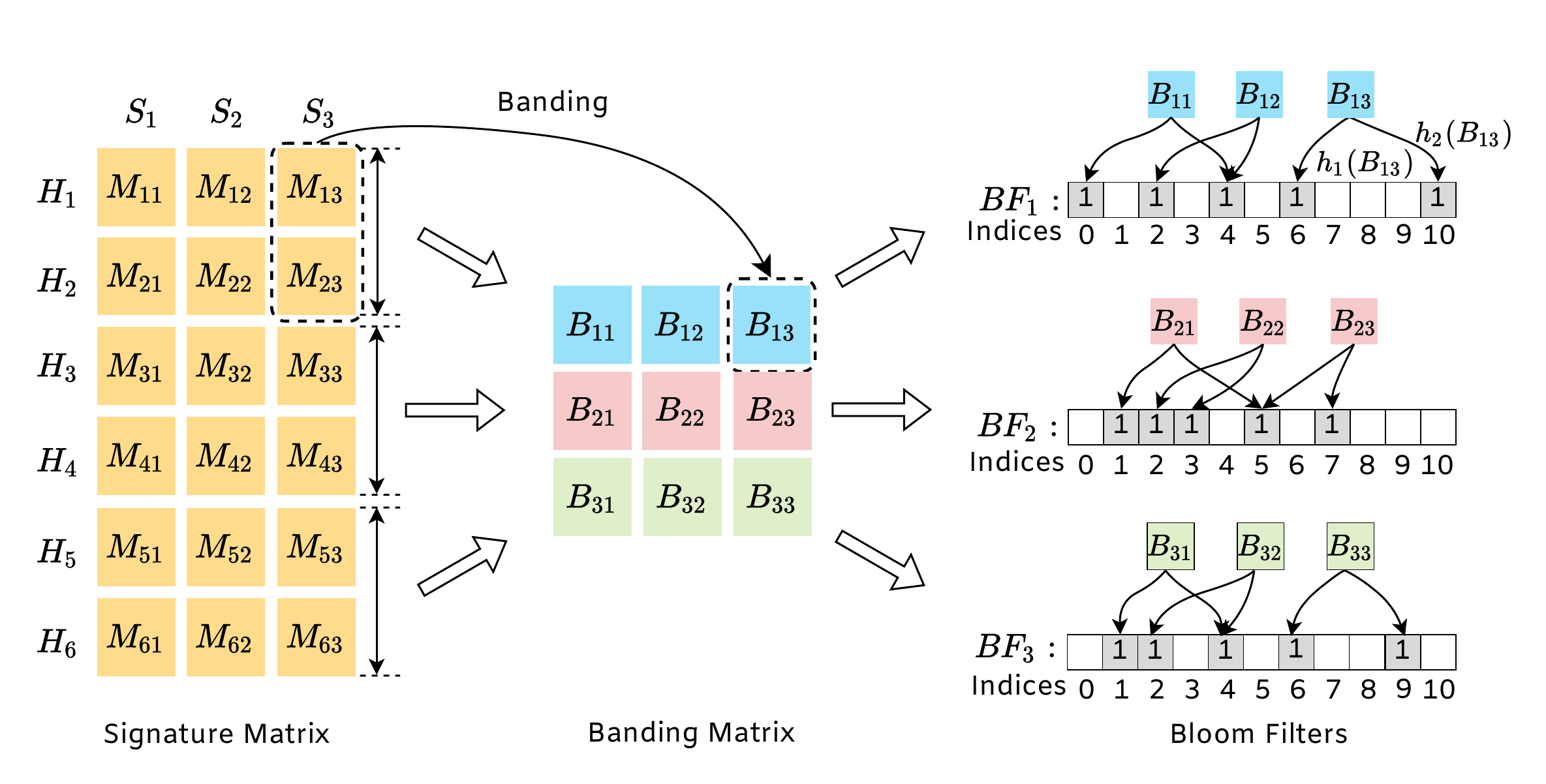}}
\caption{Example of inserting transaction sets into Bloom filters. We consider three sets $S_1$, $S_2$, and $S_3$ and 6 MinHash functions. The band width $r=2$ }
\label{fig:bloom_filter}
\end{center}
\end{figure}




\subsubsection{Inserting sets into Bloom filters.}
We adopt the MinHash algorithm (cf. Section~\ref{ssec: minhash}) as the approach to implement LSH.
Take $\mathcal{S}^p$ as an example, $\Cli_p$ first employs $m$ distinct minhash functions $H$ (cf. Equation~\ref{equ: minhash}) on each set $S_i\in \mathcal{S}_p$
resulting in a signature matrix $M_S^p$ with $m$ rows and $|\mathcal{S}^p|$ columns, where $|\cdot|$ denotes the number of sets in $\mathcal{S}_p$. Each row of the matrix represents applying the same minhash function to all sets in $\mathcal{S}_p$, and each column represents applying all minhash functions to the same set.

A banding technique then be applied to the matrix. Specifically, we divide the matrix into bands, each containing $r$ rows of the matrix, resulting in a total of $K=m/r$ bands. Each column of a band, which is composed of the result of applying  $r$ minhash functions to one set, can be treated as a result of applying LSH on the set. If two sets have the Jaccard similarity of $s$, then the probability that their columns within the same band are equal is $s^r$. By mapping each column to a distinctive signature, for example, by utilizing the MD5 function~\cite{md5}, each band can be treated as one row of the banding matrix. We denote it as $B_S^p$.
We then insert each band into a Bloom filter (cf. Section~\ref{ssec: bloomfilter}), resulting in $K$ Bloom filters $BF_S^p[1], ..., BF_S^p[K]$. When the context is clear, we omit the superscripts and
subscripts, and represent each Bloom filter as $BF_k, k \in \{1,...,K\}$.

We note that to guarantee the LSH of two similar sets are equal with high probability, $\Cli_p$ and $\Cli_q$ are required to use the same MinHash functions on $\mathcal{S}_p$ and $\mathcal{D}_q$.
Figure~\ref{fig:bloom_filter} presents an example of inserting three sets $S_1$, $S_2$, and $S_3$ into three Bloom filters, with the band with $r=2$.

\subsubsection{Probability-based similar set detection.}
\begin{figure}[t]
\begin{center}
\centerline{
    \includegraphics[width=0.85\linewidth]{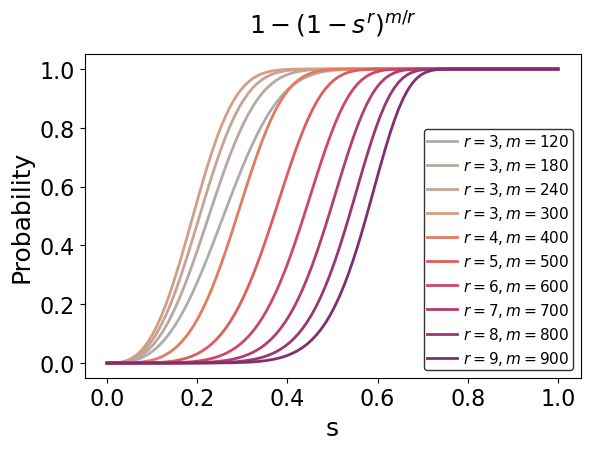}}
\caption{The probability calculated with Equation~\ref{fun:prob} with different $r$ and $m$.}
\label{fig:f_exlaination}
\end{center}
\end{figure}

With the received Bloom filters, institutions can detect whether the node is involved in money laundering activities by querying the existence of the band values in corresponding Bloom filters. Specifically, for a set $S_i \in \mathcal{S}^p$, denote its band values as $B_{i}$, which are a column in the banding matrix. $\Cli_p$ query the existence of each $B_{ki}$ in the corresponding Bloom filter $BF_k$. Theoretically, if there exists a set $D_j \in \mathcal{D}^q$ that exhibits a similarity of \(s\) with \(S_i\), the probability that at least one Bloom filter contains $B_{ki}$ is:
\begin{equation}
\label{fun:prob}
1 - (1 - s^r)^{K},    
\end{equation}
where $K=m/r$. As shown in Figure~\ref{fig:f_exlaination}, by appropriately selecting values for $m$ and $r$, this probability can be adjusted to be close to 1 or 0, depending on the level of similarity. For example, when the threshold is $0.4$, we set $r=4$ and $m=400$ so the probability is about $0.92$.
Consequently, if at least one $B_{ki}$ tested exists in $BF_k$, we treat the corresponding node of $S_i$ as a potential source within a money laundering subgroup.

With the probability-based similar set detection, we denote our AML method as \textbf{Prob-CSGM} and Algorithm~\ref{alg: IsMoneyLaunderingProb} presents the pseudo-code of the method.

\begin{algorithm}[htb]
\caption{IsSimilarSetProb}
\label{alg: IsMoneyLaunderingProb}
\small
\KwIn{$S_i$, $BF_k, k\in\{1,...,K\}$} 
\KwOut {$h$, $h=1$ indicates the set is a similar set}
\BlankLine
$h=0$\\
\For{$k \in \{1,...,K\}$}{
    $B_{ki}\leftarrow \operatorname{BANDING}(\{H_t(S)|t\in[(k-1)r, kr]\})$\\
    \If{$B_{ki} \in BF_k)$}{
    $h=1$\\
    break
    }
}
    \Return{h}

\end{algorithm}



\subsubsection{Similarity-based similar set detection} 
While the probability-based method enables the detection of source/destination nodes involved in money laundering activities, it suffers from a high false positive rate, as dissimilar sets may still be detected in at least one Bloom filter. Conversely, even when the threshold is increased, there remains a possibility that similar sets may go undetected.
To address this limitation, we propose an alternative approach to estimate the similarity directly.

Recall that each institution divides the signature matrix $M$ into $K$ bands. The probability that any two columns within a band are identical is given by $s^r$, where $s$ is the similarity we aim to estimate. A straightforward approach is to set $r = 1$ and estimating the similarity by calculating the ratio of Bloom filters that contain the band value of the transaction set to the total number of Bloom filters.
Formally, let $t_i$ be a random variable that equals 1 if $B_{ki}$ is present in the $k$-th Bloom filter $BF_k$ and 0 otherwise. The estimated Jaccard similarity is then given by $\frac{1}{m} \sum_{k=1}^{K} t_i$.
A set $S_i$ is flagged as a money laundering set if this estimated similarity exceeds a predefined threshold $\tau$.

However, applying the above method introduces significant bias in the similarity estimation. This bias occurs because a Bloom filter stores hash values from all sets 
$D \in \mathcal{D}$, and multiple sets may share overlapping elements with $S_i$. As a result, this overlap leads to an overestimation of similarity, as the Bloom filter cannot differentiate between the contributions of different sets that share elements with $S_i$.
Specifically, assume that sets $D_1, \dots, D_Q \in \mathcal{D}$ have overlapping elements with $S_i$, such that $J(S_i, D_q) > 0$ for $q \in \{1,...,Q\}$. Let $A_q$ denote the event that $B_i^S = B_q^D$. The probability of this event occurring is given by $\Pr(A_q) = \frac{|S_i \cap D_q|}{|S_i \cup D_q|}$, which is the Jaccard similarity between $S_i$ and $D_q$.
Let $z$ be a random variable that equals 1 if $B_i^S$ is detected in the Bloom filter and 0 otherwise, we have
\begin{equation}
\begin{aligned}
    Pr[z=1] &= Pr(\bigcup_{k=1}^K A_k)\\
                &=S_1-S_2+\cdots+(-1)^{n-1} S_n\\
                &\leqslant \min \left\{S_1, 1\right\}
\end{aligned}
\end{equation}
where $S_k=\sum_{1 \leqslant i_1<\cdots<i_k \leqslant n} P\left(A_{i_1} \cap \cdots \cap A_{i_k}\right)$.


To solve the problem, we propose estimating $s^r ,r\geq 1$ instead of $s$. 
The reason is that, to detect money laundering groups, we can focus on the largest similarity between $S_i$ and any set $D_q$, defined as $P := \max\{Pr(A_1), \ldots, Pr(A_q)\}$. By estimating $s^r$, the probability that less similar sets produce equivalent banding values to $S_i$ is reduced, resulting in a more accurate estimation of $P^r$, and hence $P$.
For example, consider two sets, $D_1$ and $D_2$, with similarities of $0.8$ and $0.2$ with $S_i$, respectively. Directly estimating $s$ can introduce a bias of up to $0.2$, as $Pr[z=1] - 0.8 < 0.2$. When $r=2$, the probability that the banding values are equal is $0.81$ for $D_1$ and $0.04$ for $D_2$. This results in a more accurate estimation of $P$, as $\sqrt{Pr[z=1]} - 0.8 < \sqrt{0.64+0.04} - 0.8 \simeq 0.02$. A formal analysis is presented as follow:

\begin{theorem}
    \label{thm:banding}
    Suppose that $X_1, ..., X_N$ are a sequence of real values with $0\leq  X_N \leq ... \leq X_1\leq1$. Then $\forall \varepsilon>0$, when $r>\log_p(\frac{\varepsilon}{X_1 (N-1)})$,
    $$
    (\sum_{i=1}^N X_1^r)^{1/r} - X_1 \leq \varepsilon.
    $$
\end{theorem}

The proof is presented in Appendix~\ref{ssec: theoretical_ayalysis}. Based on the analysis, we estimate $P^r$ as $l/K$, where $l = \sum_{k=1}^{M/r} \mathbb{1}[B_{ki} \in BF_k]$, which represents the number of occurrences where $B_{ki}$ is found in $BF_k$.
We identify sets involved in money laundering if their similarities exceed a predefined threshold. We can define this threshold as $\tau^r$ or estimate the similarity as $\left(l/K\right)^{1/r}$.

We denote the method as \textbf{Sim-CSGM} and present in the pseudo-code in Algorithm~\ref{alg: IsMoneyLaunderingSim}. 

\begin{algorithm}[htb]
\caption{IsSimilarSetSim}
\label{alg: IsMoneyLaunderingSim}
\small
\KwIn{Query set $S_i$, Bloom filters $BF_k,k\in\{1,..,K\}$} 
\KwOut {$h$, $h=1$ indicates the set is a similar set}
\BlankLine
\For{$k \in \{1,...,M/r\}$}{
    $B_{ki}\leftarrow \operatorname{BANDING}(\{H_t(S)|t\in[(k-1)r, kr]\})$\\

    \eIf{$B_{ki} \in BF_k$}{$t_k =1$}{$t_k =0$}
}
    $s_{i}  = (\sum_{k=1}^{K}t_k r)/m$\\
    \eIf{$s_{i}\geq \tau^r$}{h=1}{h=0}
    \Return{h}

\end{algorithm}

%% file: sections/experiments.tex
\section{Experiments}
In this section, we conduct extensive experiments to verify the effectiveness and efficiency of CSGM. 

\begin{table*}[!ht]
    \centering
    \caption{Experiments on AMLSim and AMLWorld datasets. "-" represents that the metric is unsuitable for the method. We bold the best experimental results and underline the second-best results.}
    \label{tab:exp_effectiveness}
    \resizebox{\textwidth}{!}{
    \begin{tabular}{ccccccccccc}
        \toprule  
         & \multicolumn{5}{c}{\bal} & \multicolumn{5}{c}{\unb} \\
        \cmidrule(r){2-6} \cmidrule(r){7-11} 
        \textbf{Methods} & ACC & Precision & Recall & F1-score & AUC & ACC & Precision & Recall & F1-score & AUC \\
        SGM & 0.9761 & 0.8627  & 0.9047 & 0.8832 & \textbf{0.9743} & 0.9805 & 0.8665 & 0.9678 & 0.9144 & 0.9876\\
        GIN~\cite{xu2018powerful, hu2019strategies} & 0.9497$\pm$0.0021 & 0.8397$\pm$0.0096 & 0.8992$\pm$0.0033 & 0.8684$\pm$0.0047 &0.9301$\pm$0.0016 & 0.8978$\pm$0.0064 & 0.6926$\pm$0.0159 & 0.9045$\pm$0.0034 & 0.7844$\pm$0.0109 & 0.9003$\pm$0.0049 \\ 
        GAT~\cite{velivckovic2017graph} & 0.8332$\pm$0.0123 & 0.5285$\pm$0.0201 & 0.9173$\pm$0.0021 & 0.6704$\pm$0.0159 & 0.8657$\pm$0.0071 & 0.8235$\pm$0.0181 & 0.5424$\pm$0.0288 & 0.9251$\pm$0.0035 & 0.6835$\pm$0.0225 & 0.8611$\pm$0.0113 \\
        PNA~\cite{velickovic2019deep} & 0.9533$\pm$0.0017 & 0.8508$\pm$0.0078 & 0.9061$\pm$0.0018 & 0.8776$\pm$0.0039 & 0.9351$\pm$0.0011 & 0.9177$\pm$0.0043 & 0.7470$\pm$0.0125 & 0.9066$\pm$0.0022 & 0.8190$\pm$0.0079 & 0.9136$\pm$0.0032\\
        LaundroGraph~\cite{cardoso2022laundrograph} & 0.936$\pm$0.0014 & 0.7870$\pm$0.0048 & 0.8961$\pm$0.0034 & 0.8380$\pm$0.0031 & 0.9206$\pm$0.0018 & 0.9136$\pm$0.0043 & 0.7350$\pm$0.0126 & 0.9070$\pm$0.0077 & 0.812$\pm$0.0077 & 0.9112$\pm$0.0029 \\
        MultiGIN~\cite{egressy2024provably} & 0.9827$\pm$0.0003 & \textbf{0.9949$\pm$0.001} & \uline{0.9108$\pm$0.0016} & \uline{0.9510$\pm$0.0008} & 0.9549$\pm$0.0008 & 0.9809$\pm$0.0005 & \textbf{0.9955$\pm$0.0012} & \uline{0.9110$\pm$0.0018} & \uline{0.9514$\pm$0.0012} & \uline{0.9550$\pm$0.0009}\\
        \midrule
        Prob-CSGM & \uline{0.9858$\pm$0.0007} & \uline{0.9926 $\pm$0.0003} & 0.8638$\pm$0.0072 & 0.9237$\pm$0.0041 & - &\uline{0.9880$\pm$0.0019} & 0.9928$\pm$0.0010 & 0.8943$\pm$0.0176 & 0.9409 $\pm$0.0096 & - \\
        Sim-CSGM & 
        \textbf{0.9908$\pm$0.0006} & 0.9833$\pm$0.0012 & \textbf{0.9231$\pm$0.0068} & \textbf{0.9522$\pm$0.0033} & \uline{0.9607$\pm$0.0034} & \textbf{0.9964$\pm$0.0012} & \uline{0.9930$\pm$0.0002} & \textbf{0.9737$\pm$0.0114} & \textbf{0.9833$\pm$0.0058} & \textbf{0.9865$\pm$0.0057} \\
        \bottomrule 
        \toprule
         & \multicolumn{5}{c}{\hi} & \multicolumn{5}{c}{\li}\\
         \cmidrule(r){2-6} \cmidrule(r){7-11} 
          \textbf{Methods} & ACC & Precision & Recall & F1-score & AUC & ACC & Precision & Recall & F1-score & AUC\\
         \midrule
         SGM & 0.9992 & 0.5187  & 0.6501 & 0.5770 & 0.8250 & 0.9989 & 0.0314 & 0.1765 & 0.0533 & 0.5878 \\         
         GIN~\cite{xu2018powerful, hu2019strategies} & 0.9984$\pm$0.0004 & 0.2938$\pm$0.0781 & 0.5526$\pm$0.0892 & 0.3811$\pm$0.0828 & 0.7757$\pm$0.0447 & 0.9997$\pm$0.0001 & 0.1791$\pm$0.0350 & 0.1647$\pm$0.0738 & 0.1598$\pm$0.0474 & 0.5823$\pm$0.0369\\ 
         GAT~\cite{velivckovic2017graph} & 0.9992$\pm$0.0001 & 0.5572$\pm$0.1188 & 0.2143$\pm$0.0136 & 0.3081$\pm$0.0326 & 0.6071$\pm$0.0068 & 0.998 & 0.0 & 0.0 & 0.0 & 0.5 \\
         PNA~\cite{velickovic2019deep} & 0.9985$\pm$0.0001 & 0.3565$\pm$0.0208 & \textbf{0.9380$\pm$0.0097} & 0.5165$\pm$0.0234 & \textbf{0.9683$\pm$0.0049} & 0.9997$\pm$0.0001 & 0.3321$\pm$0.0052 & \uline{0.8873$\pm$0.0069} & \uline{0.4833$\pm$0.0065} & \uline{0.9435$\pm$0.0035}\\
         LaundroGraph~\cite{cardoso2022laundrograph} & 0.9992$\pm$0.0001 & 0.5412$\pm$0.0840 & 0.6193$\pm$0.0613 & 0.5710$\pm$0.0299 & 0.8094$\pm$0.0306 & 0.9998$\pm$0.0043 & 0.3846$\pm$0.0126 & 0.0490$\pm$0.0077 & 0.0870$\pm$0.0077 & 0.5245$\pm$0.0029 \\
         MultiGIN~\cite{egressy2024provably} & 0.9996$\pm$0.0002 & 0.6945$\pm$0.0959 & \uline{0.9366$\pm$0.0173} & \uline{0.7943$\pm$0.0658} & \uline{0.9681$\pm$0.0086} & 0.9996$\pm$0.0001 & 0.1746$\pm$0.0365 & 0.3353$\pm$0.2252 & 0.2104$\pm$0.1101 & 0.6675$\pm$0.1125\\
         \midrule
         Prob-CSGM & \uline{0.9996$\pm$0.0001} & \textbf{0.8747$\pm$0.0242} & 0.6413$\pm$0.0643 & 0.7392 $\pm$0.0499 & - &
        \uline{0.9998$\pm$0.0001} & \uline{0.4370$\pm$0.0684} & 0.3529$\pm$0.1038 & 0.3878 $\pm$0.086 & - \\
        Sim-CSGM & \textbf{0.9997$\pm$0.0001} & \uline{0.7718$\pm$0.0191} & 0.8292$\pm$0.0136 & \textbf{0.7995$\pm$0.0128} & 0.9145$\pm$0.0068& \textbf{0.9999 $\pm$0.0009} & \textbf{0.6458$\pm$0.0008} & \textbf{0.9118$\pm$0.0001} & \textbf{0.7561$\pm$0.0041} & \textbf{0.9558$\pm$0.0002} \\
        \bottomrule 
    \end{tabular}}
\end{table*}

\begin{figure*}
\centering
\begin{tabular}{cccc}
    \subfloat{
        \includegraphics[width=0.23\textwidth,valign=c]{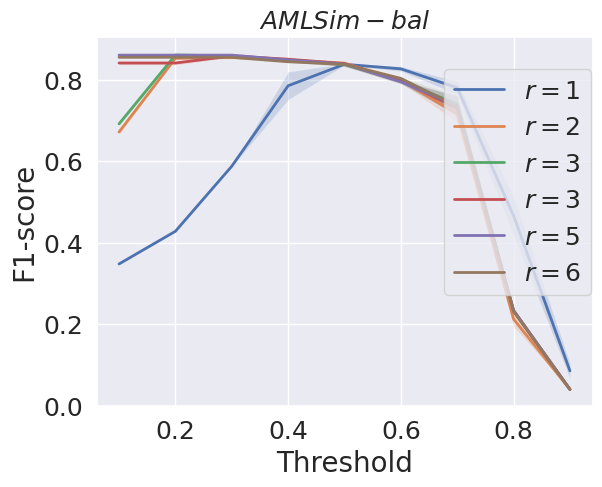}} &
    \subfloat{
        \includegraphics[width=0.23\textwidth,valign=c]{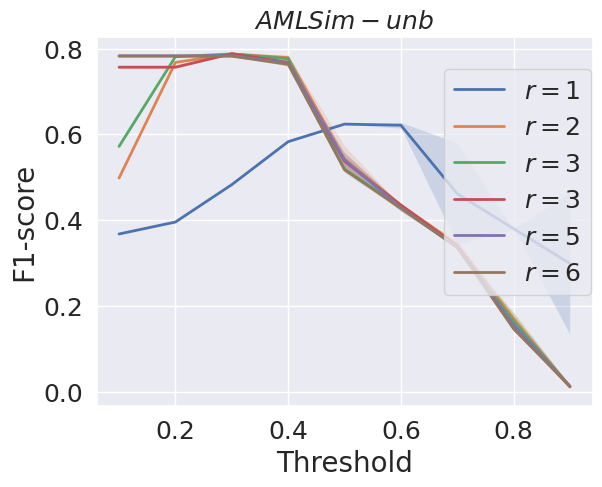}} &
    \subfloat{
        \includegraphics[width=0.23\textwidth,valign=c]{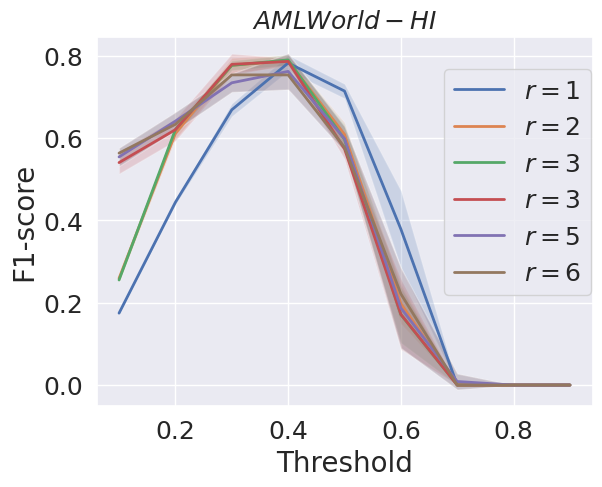}} &
    \subfloat{
        \includegraphics[width=0.23\textwidth,valign=c]{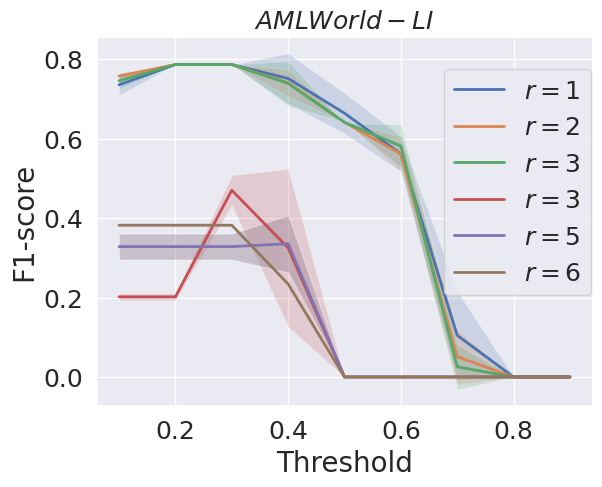}}\\
\end{tabular}
\caption{Experiments for the similarity-based method with the four synthetic datasets.}
\label{fig:sim_thd}
\end{figure*}

\subsection{Experimental setting}

\Paragraph{Dataset.}
To evaluate whether CSGM can detect money laundering activities in practice, we construct a real-world dataset called \textit{Alipay-ECB}, based on daily transactions recorded on Alipay~\cite{alipay} and E-Commerce Bank~\cite{ECB}. It comprises over 200 million accounts and 300 million transactions.
To the best of our knowledge, it is the largest transaction dataset that tracks currency flow in the real world, providing a comprehensive reflection of money laundering activities. Detailed information on the dataset is provided in Appendix~\ref{ssec:dataset}.
We also conduct experiments on synthetic datasets that simulate transactions for money laundering activities.
We utilize AMLSim~\cite{AMLSim} and AMLWorld~\cite{altman2024realistic}, which supports building a multi-agent simulator of anti-money laundering and has been widely used in previous works~\cite{karim2023catch,weber2018scalable,usman2023intelligent,egressy2024provably}.
For AMLSim~\cite{AMLSim}, we synthesize two datasets with \numprint{100000} nodes, and simulate two scenarios where two institutions have balanced transaction subgraphs or unbalanced subgraphs. We refer to $\textit{AMLSim\_bal}$ and $\textit{AMLSim\_unb}$, respectively. For AMLWorld~\cite{altman2024realistic}, we choose two datasets of size 5 million and 7 million for experiments. We refer to $\textit{AMLWorld\_HI}$ and $\textit{AMLWorld\_LI}$, where HI stands for relatively higher illicit rate and LI stands for lower illicit rate.
The statistics of the datasets are presented in Table~\ref{tab: datasets}.

\Paragraph{Baselines.}
\textbf{SGM} refers to the centralized scatter-gather mining method described in Section~\ref{ssec: MLSD}.
A line of research leverages graph neural networks (GNNs) for identifying money laundering transactions. \textbf{GIN}\cite{xu2018powerful, hu2019strategies}, \textbf{GAT}\cite{velivckovic2017graph}, and \textbf{PNA}\cite{velickovic2019deep} are commonly used GNN models for general graph classification tasks. Two additional studies propose GNNs specifically tailored for money laundering detection. \textbf{LaundroGraph}\cite{cardoso2022laundrograph} introduces a self-supervised graph representation learning method to detect money laundering. \textbf{MultiGIN}\cite{egressy2024provably}, incorporates a range of adaptations, including multigraph port numbering, ego IDs, and reverse message passing, to enhance GNNs' ability to detect various patterns of illicit activities. 

\newenvironment{packeditemize}{
\begin{list}{$\bullet$}{
\setlength{\labelwidth}{8pt}
\setlength{\itemsep}{0pt}
\setlength{\leftmargin}{\labelwidth}
\addtolength{\leftmargin}{\labelsep}
\setlength{\parindent}{0pt}
\setlength{\listparindent}{\parindent}
\setlength{\parsep}{0pt}
\setlength{\topsep}{3pt}}}{\end{list}}


\subsection{Experiments on Alipay-ECB.}
\Paragraph{Data process.}
We first preprocess the dataset, including account segregation, transaction aggregation, and transaction filtering to remove noise transactions from the dataset. As a result, we obtain a transaction graph with 48.95 million accounts and 34.45 million transactions. The detailed process is presented in Appendix~\ref{ssec:dataset}.

\Paragraph{Results.}
We experiment with similarity-based methods and set the threshold to $0.1$.
Examples of detected groups are presented in Figure~\ref{fig:detected_groups}.
We randomly selected 100 detected groups and evaluated them based on whether the accounts in each group had been reported as illegal~\footnote{For reasons related to corporate confidentiality and data safety, we regret that we cannot disclose the exact figures of detected groups.}.

Among the 100 detected groups, 59 were identified as money laundering groups, with more than half of their accounts recognized as illicit in actual business operations. Two groups were identified as non-money laundering, while the remaining 39 groups contained at least one illicit account.
The results demonstrate that our method can effectively detect money laundering groups. With the group-level information, our methods would assist in uncovering previously undiscovered illicit accounts.

\subsection{Experiments on synthetic datasets.}
We experiment with Prob-CSGM and Sim-CSGM on four synthetic datasets and compare them with all five baselines. 
We set the number of hash functions $m$ used in MinHash to $100$, and $r$ to $5$ for Prob-CSGM. 
For Sim-CSGM, $m$ is set to $100$, $r=2$, and the threshold is $0.2$ for $AMLSim$ and $0.3$ for $AMLWorld$.
The size of the Bloom filter is 500,000 bits for AMLSim and 3,000,000 bits for AMLWorld, resulting in a false positive probability of approximately $0.01$.

The results are shown in Table~\ref{tab:exp_effectiveness}. On the two AMLSim datasets, our methods have a performance comparable to SGM. Sim-CSGM, in particular, outperforms the centralized method in terms of recall, indicating that it is more effective at identifying abnormal nodes comprehensively. On the AMLWorld datasets, SGM exhibits low precision, with even poorer performance on AMLWorld-LI. This is because SGM tends to identify small transaction groups as money laundering groups, which is normal in AMLWorld-HI. However, Sim-CSGM can still identify money laundering groups, demonstrating its robustness.

Compared to GNN-based methods, our approach has a comparable performance on the AMLSim dataset, which achieves the best results with both precision and recall rates exceeding 90\%. However, when the proportion of illicit transactions is exceptionally low, such as in the AMLWorld-HI dataset, MultiGIN suffers from low precision, leading to a high false positive rate. In contrast, our methods maintain strong performance on the AMLWorld datasets, highlighting the generalizability of our approach.

\subsection{Ablation study.}
To explore the impact of different parameters on the performance of our methods, we conducted experiments by varying the number of hash functions used in MinHash $m$, the number of rows $r$ in each band as well as the threshold.


Here, we mainly focus on Sim-CSGM. 
We vary the threshold from 0.2 to 0.6 and observe the change of F1-score with different $r$.
The results are depicted in Figure~\ref{fig:sim_thd}.
It shows that the F1-score when $r>1$ performs better than when $r=1$, showing the effectiveness of the banding technique in the similarity-based method. Furthermore, when $r=1$, the method prefers a higher threshold, illustrating that repeated elements in a band lead to an overestimation of similarity when using the bloom filter. 

Additionally, experiments in Appendix~\ref{ssec:rows_r} show that the banding technique could significantly reduce the number of repeated elements. We also evaluate the efficiency of our methods in terms of the communication costs as well as the running time in Appendix~\ref{ssec:efficiency}. The results show that our methods take only a few minutes.


%% file: sections/related_works.tex
\section{Related works}
The term money laundering was first used at the beginning of the 20th Century to label the operations that in some way intended to legalize the income derived from illicit activity,  thus facilitating their entry into the monetary flow of the economy~\cite{AMLDef}. Since then, numerous methods have been proposed to identify money laundering activities~\cite{hooi2016fraudar,le2010application,michalak2011graph,rajput2014ontology,soltani2016new,starnini2021smurf,zhang2003applying}.  
Rule-based approaches were first widely used in the early days~\cite{michalak2011graph,rajput2014ontology}. Rajput et al.~\cite{rajput2014ontology} propose an ontology-based expert system to detect suspicious transactions, and Michalak et al.~\cite{michalak2011graph} propose a method that integrates the fuzzing method and decision rules to detect suspicious transactions. 
Although easy to deploy, rule-based methods can easily be evaded by fraudsters.

With the popularity of machine learning, learning-based methods have become an emergency. Tang et al.~\cite{tang2005developing} propose to use the support vector machine method (SVM) to detect unusual behaviors in transactions. 
Lv et al.~\cite{lv2008rbf} judge whether the capital flow is involved in money laundering activities using RBF neural networks calculated from time to time. Paula et al.~\cite{paula2016deep} also show some success for AML by using deep neural networks. However, these methods detect money laundering activities in a supervised manner, suffering from highly skewed labels and limited adaptability.

Graphs have the advantage of better characterizing the association between objects. Many graph-based anomaly detection techniques have been developed for discovering structural anomalies. 
Zhang et al.~\cite{zhang2003applying} use financial transaction networks and community detection algorithms to find money laundering groups.
Cardoso et al.~\cite{cardoso2022laundrograph} introduces a self-supervised graph representation learning method aimed at detecting money laundering. Recently, Béni et al.~\cite{egressy2024provably}, incorporates a range of adaptations, including multigraph port numbering, ego IDs, and reverse message passing, to enhance GNNs' ability to detect various patterns of illicit activities. 

Despite the advance of all those methods, they work based on the prerequisite that the transaction graph is centralized, while in practice, money laundering activities span across multiple institutions s.t. the transaction subgraph within one institution appears to be normal.
Our methods make the \textit{first} steps towards collaborative anti-money laundering among institutions without exposing the transaction graphs.

%% file: sections/conclusion.tex
\section{Conclusion}
In this work, we propose the \textit{first} algorithm enabling collaborative anti-money laundering (AML) among institutions while preserving the privacy of their transaction graphs. We employ LSH~\cite{lsh} and bloom filters~\cite{bloomfilter} to reduce communication costs and enhance efficiency. Experimental evaluations on two synthetic datasets demonstrate the effectiveness and efficiency of the proposed algorithm. In future work, we will attempt to deploy the algorithm in real-world industrial settings to evaluate its effectiveness with realistic data. Moreover, we will enhance the algorithm to address intricate money laundering scenarios involving more institutions and more complex transaction graphs.

%% file: sections/appendix.tex
\section{Appendix}


\begin{table}[hb]
\small
\centering
\caption{Summary of notations}
\begin{tabular*}{8cm}{r|l}
\toprule
\textbf{Notation} & \textbf{Description} \\ 
\midrule
$\mathcal{G}$ & transaction graph\\
$\mathcal{G}'$ & transaction graph with inversed direction \\
$m$ & number of hash functions used in MinHash\\
$r$ & number of rows of each band \\ 
$K$ & number of bloom filters \\ 
$S$ & the set of cross-institution transactions find with $\mathcal{G}$ \\ 
$D$ & the set of cross-institution transactions find with $\mathcal{G}'$ \\
$\mathcal{S, D}$ & list of sets $S$ and $D$ \\
$M$ & Signature matrix\\
$B$ & Banding matrix\\

\bottomrule
\end{tabular*}
\label{tab:notations}
\vspace{-3mm}
\end{table}

\subsection{Theoretical Analysis}
\label{ssec: theoretical_ayalysis}
\begin{customthm}{1}[restated]
    Suppose that $X_1, ..., X_N$ are a sequence of real values with $0\leq  X_N \leq ... \leq X_1\leq1$. Then $\forall \varepsilon>0$, there $\exists \delta$, s.t. when $r>\delta$,
    $$
    (\sum_{i=1}^N X_1^r)^{1/r} - X_1 \leq \varepsilon.
    $$
\end{customthm}

\begin{proof}
We have
$$
\begin{aligned}
&(\sum_{i=1}^N  X_1^r)^{1/r} - X_1\\
=& \left(x_1^r\left(1+\sum_{i=1}^{N-1}\left(\frac{X_i}{X_1}\right)^r\right)\right)^{1 / r} -X_1\\
=&X_1\left(1+\sum_{i=1}^{N-1}\left(\frac{X_i}{X_1}\right)^r\right)^{1 / r} - X_1\\
=&X_1 \left(\left(1+\sum_{i=1}^{N-1}\left(\frac{X_i}{X_1}\right)^r\right)^{1 / r} -1\right)
\end{aligned}
$$
To prove the inequality, we only need to prove 
$$
\left(1+\sum_{i=1}^{N-1}\left(\frac{X_i}{X_1}\right)^r\right)^{1 / r}  < \frac{\varepsilon}{X_1} + 1
$$
It is easy to prove that 
$$
\begin{aligned}
\left(1+\sum_{i=1}^{N-1}\left(\frac{X_i}{X_1}\right)^r\right)^{1 / r}
\leq & \left(1+ (N-1)\left(\frac{X_2}{X_1}\right)^r\right)^{1 / r}
\\< & 1+ (N-1)\left(\frac{X_2}{X_1}\right)^r
\end{aligned}
$$
When $r > \log_p(\frac{\varepsilon}{X_1 (N-1)})$, we have
$$
1+ (N-1)\left(\frac{X_2}{X_1}\right)^r < \frac{\varepsilon}{X_1} + 1,
$$
where $p = X2/X1$

\end{proof}

\subsection{Dataset Statistics}
\label{ssec:dataset}

\Paragraph{Alipay-ECB.}
Alipay Mobile Payment~\cite{alipay} is the world's largest mobile payment platform, allowing users to pay for a wide range of daily needs, including money transfers, online shopping, salary deposits, investments, and more. 
As of June 2020, Alipay serves over 1.3 billion users and 80 million merchants~\cite{alipay_wiki}, making it an invaluable resource for studying money laundering activities that may be concealed within its vast volume of transaction records. 
Meanwhile, E-Commerce Bank, operating entirely online, serves tens of millions of users and merchants across China, facilitating numerous financial transactions between Alipay and E-Commerce Bank daily. The AlipayECB dataset captures these transactions, with the majority of records originating from Alipay. These include transactions between Alipay users as well as between users and various bank accounts, with ESB being one of the many banks involved. 

\Paragraph{Time span.}
The AlipayECB dataset is constructed using transactions that occurred on Alipay and E-Commerce Bank (ECB) within a single day. Unlike synthetic datasets such as AMLSim~\cite{AMLSim} and AMLWorld~\cite{altman2024realistic}, where transactions within a single money laundering group can span several days~\cite{amlworld_kaggle}, money laundering transactions on digital platforms tend to occur rapidly. Funds are moved in and out quickly to minimize the risk of losses due to account monitoring and censorship. Based on this observation, we focus exclusively on transactions occurring within a single day.

\Paragraph{Data processing.}
To facilitate the experiments on the dataset, we process the data as follows:
\begin{packeditemize}
   \item \textbf{Account Segregation:} A user may link deposits or credit cards from different banks to their Alipay account. As a result, numerous transactions occur between Alipay and the user's cards (e.g., through withdrawal services). To facilitate the tracing of fund flows between different banks, we treat each card as a separate account, even if they belong to the same user.

   \item \textbf{Transaction Aggregation:} Transactions between two accounts may occur multiple times. However, as we mainly focus on a set of transactions, we consolidate these transactions into a single entry. This is different from MultiGIN~\cite{egressy2024provably}, which treats transactions between the same accounts as distinct entities.

   \item \textbf{Transaction Filtering:} Given that money laundering often involves large sums, we filter out transactions with small amounts (around ¥100 in our experiments).
\end{packeditemize}
After applying these steps, we obtain a transaction graph with 48.95 million accounts and 34.45 million transactions. Detailed statistics are presented in Table~\ref{tab:alipay_ecb_datasets}.

\begin{table}[hbt]
    \centering
    \caption{Statistics of Alipay-ECB after processing.}
    \label{tab:alipay_ecb_datasets}
    \resizebox{\linewidth}{!}{
    \begin{tabular}{ccccccc}
        \toprule
        \multicolumn{3}{c}{Accounts} & \multicolumn{4}{c}{Transactions}\\
        \cmidrule(r){1-3}  \cmidrule(r){4-7}
        Alipay & ECB & Others & Alipay $\rightarrow$ Alipay & ECB $\rightarrow$ ECB & Alipay$\rightarrow$ ECB & ECB$\rightarrow$ Alipay\\
        \midrule
        23.46M &  3.99M & 21.50M & 30.84M & 5.51M & 0.25M & 1.65M\\
        \bottomrule
    \end{tabular}
    }
\end{table}

\begin{figure*}[!tbp]
\centering
\begin{tabular}{ccc}
    \subfloat{
        \includegraphics[width=0.32\textwidth,valign=c]{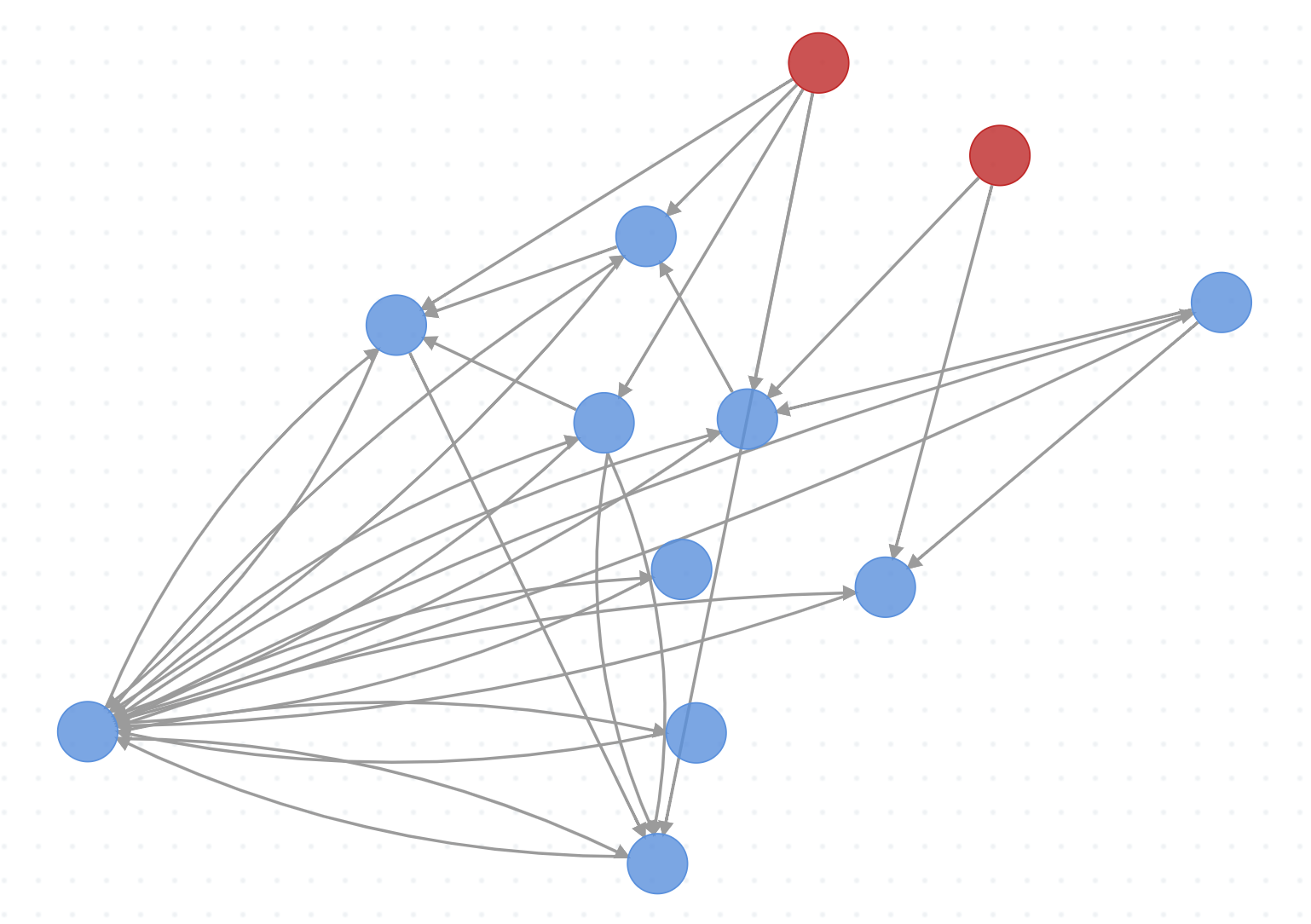}} &
    \subfloat{
        \includegraphics[width=0.33\textwidth,valign=c]{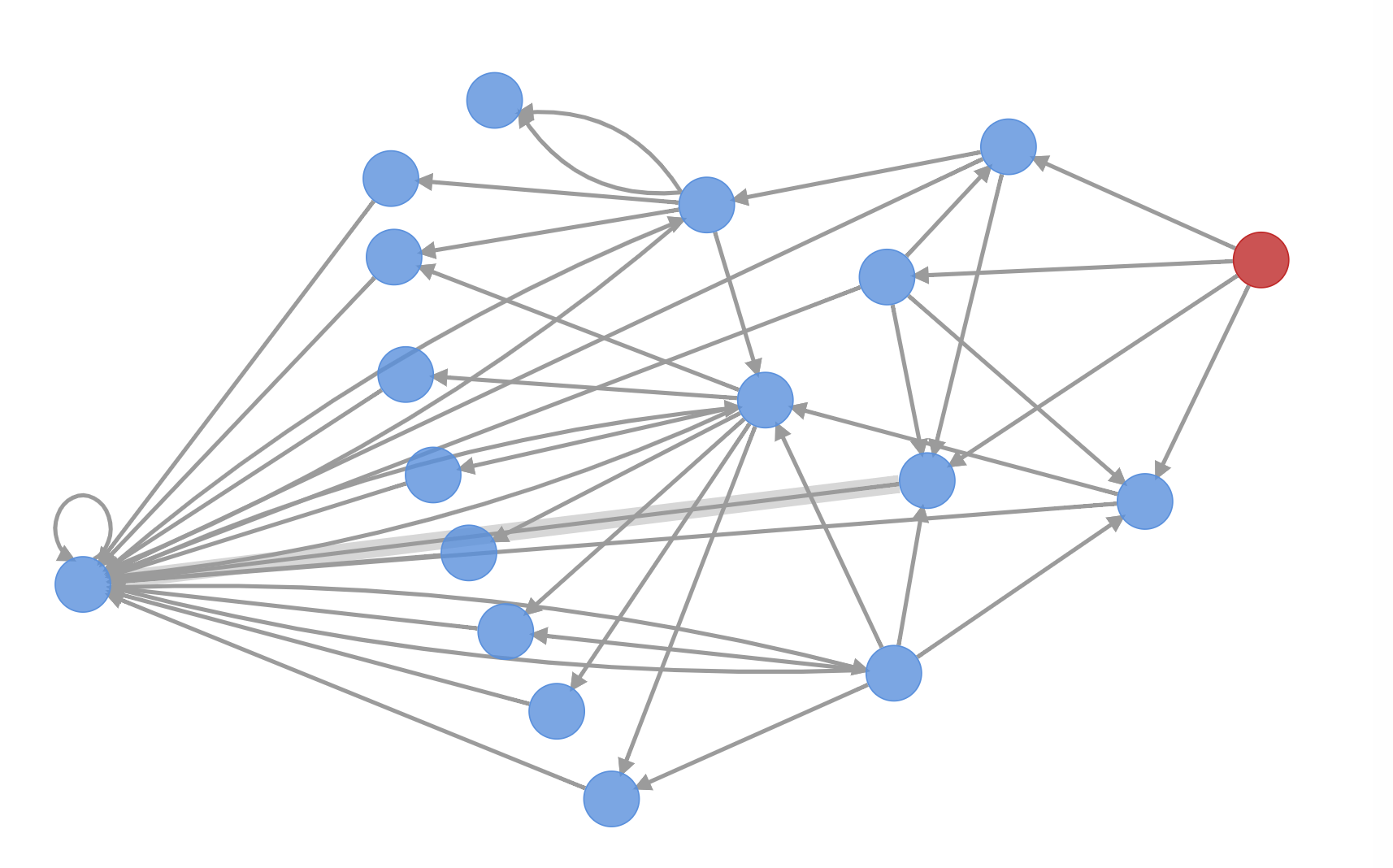}} &
    \subfloat{
        \includegraphics[width=0.33\textwidth,valign=c]{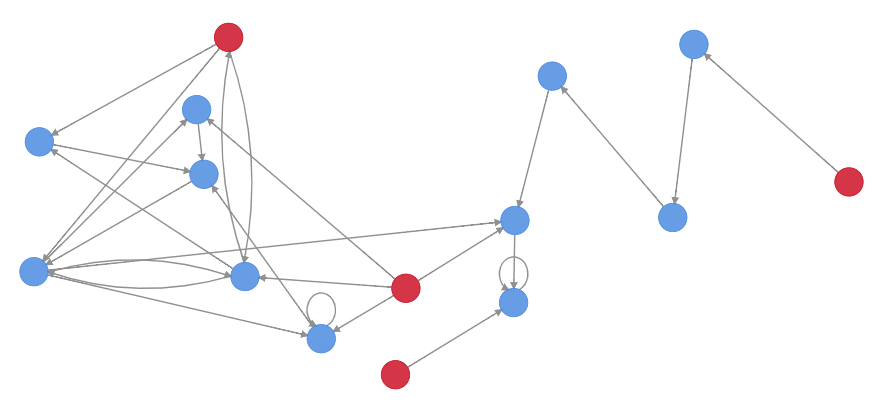}} \\
    \multicolumn{3}{c}{\textbf{(a) Money laundering groups}}\\
    \subfloat{
        \includegraphics[width=0.32\textwidth,valign=c]{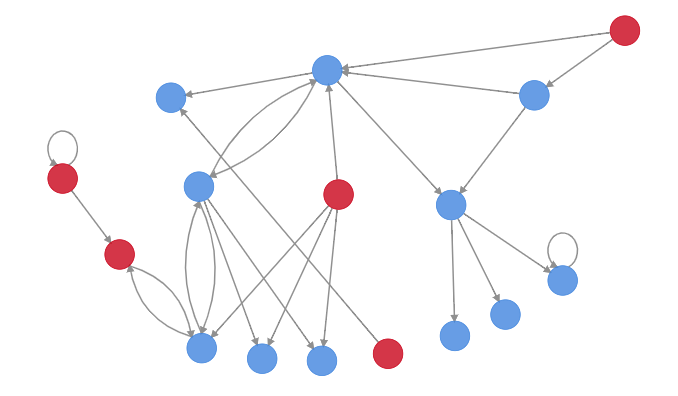}}&
    \subfloat{
        \includegraphics[width=0.33
        \textwidth,valign=c]{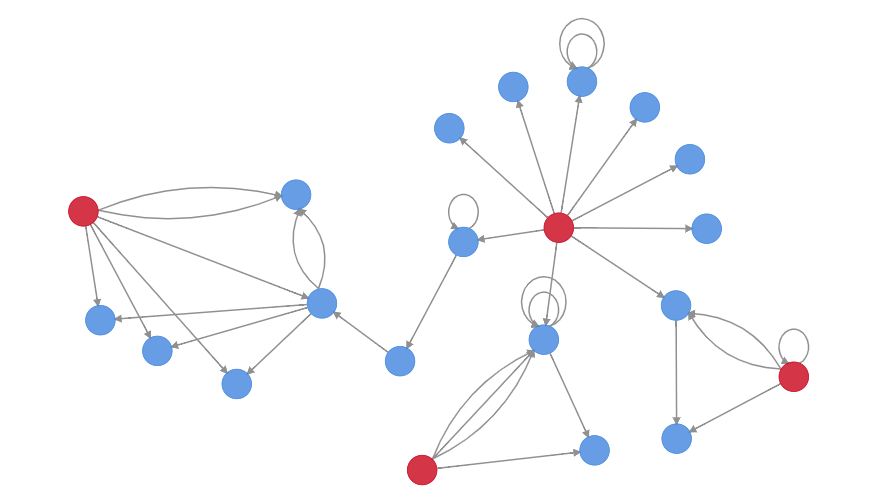}}&
    \subfloat{
        \includegraphics[width=0.32\textwidth,valign=c]{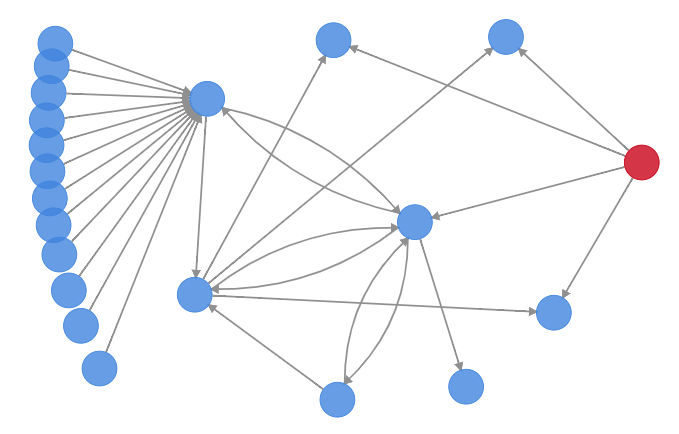}} \\
    \multicolumn{2}{c}{\textbf{(b) Grey groups}} & \textbf{(c) Normal groups}
        
\end{tabular}
\caption{Examples of money laundering groups detected by CSGM, with colors indicating different institutions. The blue nodes represent accounts from Alipay and the red nodes represent accounts from ECB.}
\label{fig:detected_groups}
\end{figure*}

\Paragraph{Examples of discovered subgraphs}
Figure~\ref{fig:detected_groups} presents examples of detected groups. Money laundering groups are identified when more than half of their accounts are classified as illicit. Grey groups are those in which only a small number of accounts have been reported for suspected money laundering activities. The normal group is associated with school financial collections.


\begin{table*}[!hbt]
    \centering
    \caption{Statistics of datasets. 
    $|\mathcal{V}|$ denotes the total number of accounts, $|\mathcal{E}|$ denote the total number of transactions, and $|\mathcal{V}_p|$ as well as $|\mathcal{E}_p|$, where $p\in[A,B]$, represent the number of accounts and transactions owned by $\Cli_p$. IR represents an illicit ratio of accounts.}
    \label{tab: datasets}
    \begin{tabular}{cccccccc}
        \toprule
        Dataset & $|\mathcal{V}|$ & $|\mathcal{E}|$&  $|\mathcal{V}_A|$ & $|\mathcal{E}_A|$ & $|\mathcal{V}_B|$ & $|\mathcal{E}_B|$ & IR\\
        \midrule
        $\textit{AMLSim\_bal}$ &  100K & \numprint{1957005} & 50K & \numprint{1653870} & 50K & \numprint{1364455} & 9.97\%\\
        $\textit{AMLSim\_unb}$ & 100K & \numprint{1959234} & 75K & \numprint{1844664} & 25K & \numprint{919211} & 10.75\%\\
        \midrule
        $\textit{AMLWorld\_HI}$ & \numprint{515020} & \numprint{5073772} & \numprint{257799} & \numprint{3718584} & \numprint{257221}& \numprint{ 3592358} & 0.070\%\\
        $\textit{AMLWorld\_LI}$ & \numprint{705861} & \numprint{6920656} & \numprint{352189} & \numprint{5096798} & \numprint{353672} & \numprint{4880541} & 0.014\%\\
        \midrule
        $\textit{Alipay-ECB}$ & 245M & 336M & 225M & 309M & 19M & 26M & -\\
        \bottomrule
    \end{tabular}
\end{table*}

\Paragraph{Machine Specs and code.}
The experiments on synthetic datasets were conducted on a single machine using Python 3. The experiments on Alipay-ECB were carried out in Java on a cluster of 20 machines.

\subsection{The impact of number of rows $r$}
\label{ssec:rows_r}
\begin{figure}[hbp]
\centering
\includegraphics[width=0.4\textwidth,valign=c]{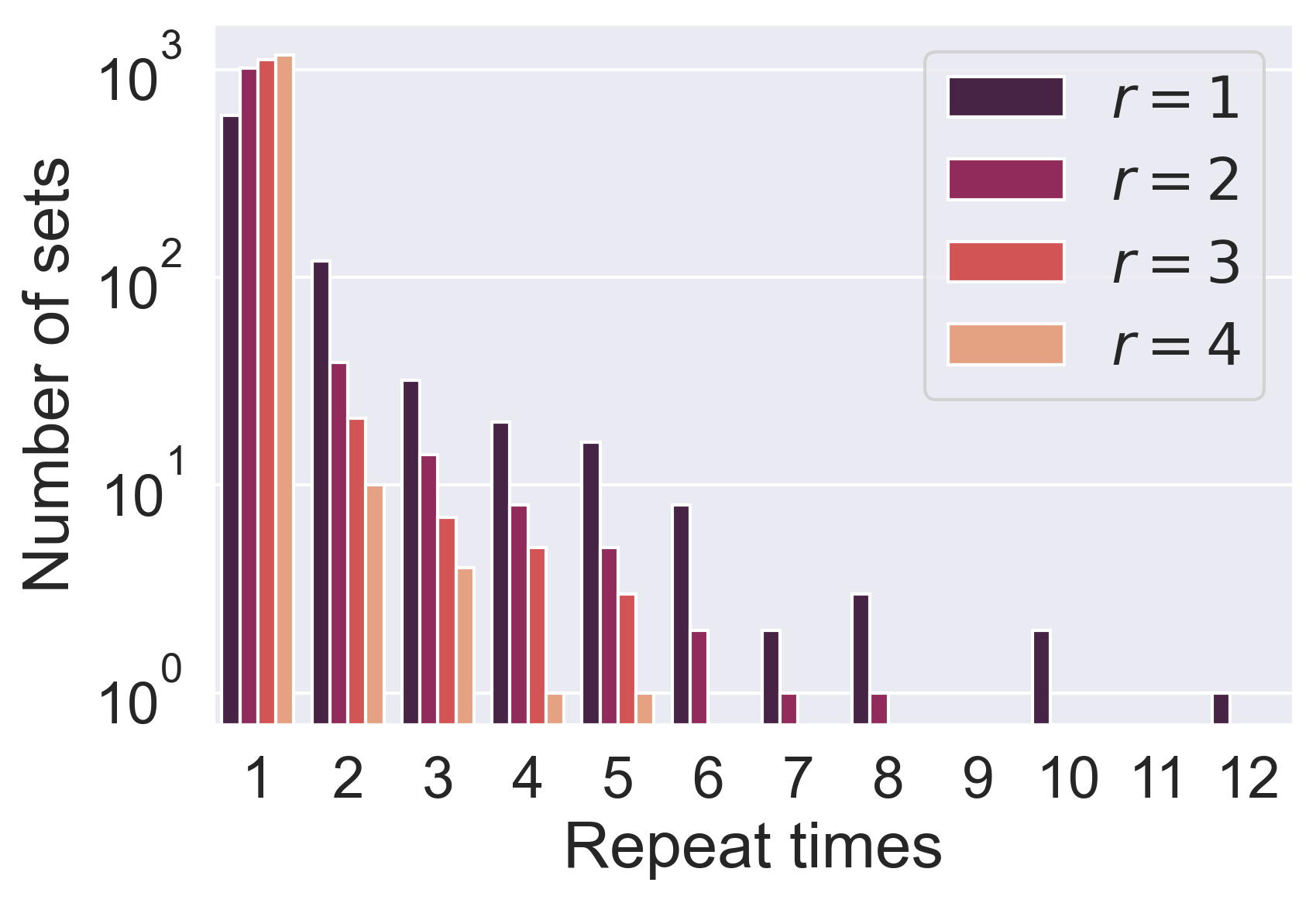}
\caption{
The frequency histogram of repeated elements in a band.}
\label{fig:sim_fix_hash}
\end{figure}

To further investigate the impact of $r$ on estimating similarity in Sim-CSGM, we plot the frequency histogram of repeated elements within a band, as shown in Figure~\ref{fig:sim_fix_hash}. The x-axis represents the number of repetitions, while the y-axis indicates the number of sets that share repeated elements with others. As the number of repeated sets increases, more bias is introduced into the similarity estimation. The results show that most sets differ from others (repeat = 1), but many still share repeated elements. However, the banding technique can significantly reduce the number of repeated elements.


\subsection{Efficiency.}
\label{ssec:efficiency}
We evaluate the efficiency of our algorithm in terms of the communication costs among institutions as well as the running time.

\Paragraph{Communication costs.}
The communication costs are primarily induced by the transmission of Bloom filters between institutions. Specifically, each institution needs to transmit its Bloom filters to another institution, with each filter containing values within a band. In our experiments, the Bloom filter size is 500,000 for AMLSim and 3,000,000 for AMLWorld, corresponding to approximately 61.04 KiB and 366.21 KiB, respectively. In our experiments, we adopt 100 Bloom filters, resulting in about 13 MiB and 75 MiB of data transfer per institution.

\Paragraph{Running time.}

We report the running time of applying our methods on the two datasets in Table~\ref{tab:running_time}. The running time for each stage during the execution of the methods is presented, including discovering sets of cross-institution transactions, performing MinHash functions on all sets, inserting bands into Bloom filters, and testing the existence of elements in Bloom filters. On the two AMLSim datasets, our methods take only a few seconds. Even on the larger AMLWorld datasets, CSGM still takes only a few minutes, demonstrating its efficiency. The MinHash calculation process is the most time-consuming, while the membership testing stage is highly efficient, requiring just a few seconds.

\begin{table*}[!htbp]
    \centering
    \caption{Running time in seconds of Sim-CSGM. \textbf{Set discovery} represents the process of discovering sets of cross-institution transactions, \textbf{minhash} presents performing minhash functions to all sets, \textbf{inserting} represents inserting bands to bloom filters, and \textbf{membership testing} represents testing the existence of elements in bloom filters..}
    \label{tab:running_time}
    \begin{tabular}{c|ccccc|c}
        \toprule  
        Dataset  & Inst. & \makecell{Set\\ Discovery} & Minhash & Inserting & \makecell{Membership\\ Testing} & Total\\
        \midrule
        \multirow{2}{*}{$\textit{AMLSim-bal}$}
        &$\Cli_A$ & 1.74\textbf{} & 3.49 & 1.80 & 0.41 & 7.44   \\
        &$\Cli_B$ & 2.26 & 7.78 & 2.10 & 0.63 & 12.77 \\
        \midrule
        \multirow{2}{*}{$\textit{AMLSim-unb}$}
        &$\Cli_A$ & 1.11 & 5.80 & 5.46 & 0.96 &  13.33 \\
        &$\Cli_B$ & 4.23 & 4.83 & 2.84 & 0.82 &  12.72   \\
        \midrule
        \multirow{2}{*}{$\textit{AMLWorld-HI}$}
        &$\Cli_A$ & 7.67 &186.45 & 10.57 & 2.93 &  207.62 \\
        &$\Cli_B$ & 1.27 & 8.70 & 6.91 & 2.56 & 19.44 \\
        \midrule
        \multirow{2}{*}{$\textit{AMLWorld-LI}$}
        &$\Cli_A$ & 7.81 &184.1 & 10.71 & 2.91 & 205.53  \\
        &$\Cli_B$ & 1.29 & 8.90 & 6.99 & 2.52 & 19.70  \\
        
        \bottomrule  
    \end{tabular}
\end{table*}





\subsection{Additional Algorithms}
\label{ssec: algs_prots}

\renewcommand*{\algorithmcfname}{Algorithm}
\begin{algorithm}[thb]
\caption{Cross-institution Transaction Set Discovery}
\label{alg:set_discovery}
\small
\KwIn{node $i$, directed graph $\mathcal{G = (V,E,X)}$, where $\vec{c}\in \mathcal{X}$ represents whether the transaction is an cross-institution transaction, depth $K$} 
\KwOut {Set of cross-institution transactions that scattered from $i$}
\BlankLine
Initialize set $S= \emptyset$\\
Denote $\mathcal{N}_i = \{j| (i, j)\in\mathcal{E}\}$, $\forall i \in \mathcal{V}$\\
Initialize set $\mathcal{Q}=\{i\}$\\
\For{$k=\{1,...,K\}$}{
    Initialize set $Z=\emptyset$\\
    \For{$v\in\mathcal{Q}$}{
        \For{$j\in\mathcal{N}_v$}{
            \eIf{$c_{v\rightarrow j} = 1$}{$S$.append($(v, j)$)}
             {$\mathcal{Z}$.append(j)}
        }
    }
    \eIf{$\mathcal{Z}==\emptyset$} {break}
    {$\mathcal{Q}=\mathcal{Z}$}
}    
\Return{$S$}
\end{algorithm}

